\documentclass[a4paper,12pt]{article}
\usepackage{amsfonts}
\usepackage{amssymb}

\usepackage[english]{babel}
\usepackage[verbose=true,twoside,hmarginratio=3:4]{geometry}
\usepackage{fabdoc}
\newtheorem{lemma}{Lemma}[section]
\newtheorem{theorem}{Theorem}[section]

\newcommand{\be}{\begin{equation}}
\newcommand{\ee}{\end{equation}}
\newcommand{\bea}{\begin{eqnarray}}
\newcommand{\eea}{\end{eqnarray}}
\newcommand{\sh}{\sinh}
\newcommand{\ch}{\cosh}

\newcommand{\calL}{{\cal L}}
\begin{document}

\title{Renormalization of Orientable Non-Commutative Complex $\Phi^6_3$ Model}

\author{Zhituo Wang\footnote{e-mail: wzht@mail.ustc.edu.cn}\, ,
  Shaolong Wan\footnote{e-mail: slwan@ustc.edu.cn}\\
Institute for Theoretical Physics and Department of Modern Physics \\
University of Science and Technology of China,\\
 Hefei, 230026, {\bf
P. R. China}}

\maketitle

\vskip 5ex

\begin{abstract}
In this paper we prove that the Grosse-Wulkenhaar type
non-commutative  orientable complex scalar $\varphi^6_3$ theory,
with two non-commutative coordinates and the third one commuting
with the other two, is renormalizable to all orders in perturbation
theory. Our proof relies on a multiscale analysis in $x$ space.
\end{abstract}

\section{Introduction}
\setcounter{equation}{0}

Since the rebirth of non-commutative quantum field theory
\cite{a.connes98:noncom,Seiberg:1999vs,DouNe,szab} , people
encountered a major difficulty. A new kind of divergences appeared
in non-commutative field theory \cite{Minw}, the UV/IR mixing. It is
a kind of infrared divergence which appears after integrating the
high scale variables and can't be eliminated. It lead people to
declare such theories non-renormalizable. But a real breakthrough of
that deadlock came from H.Grosse and R.Wulkenhaar
\cite{GrWu03-1,GrWu04-3}. They found that the right propagator for
the scalar field theory in non-commutative space should be modified
to obey the Langmann-Szabo duality \cite{LaSz}. In a series of paper
they proved that the $\varphi^4$ scalar field theory in 4
dimensional Moyal plane, $\varphi^{\star 4}_4$ for short, is
renormalizable to all orders using Polchinski's equation
\cite{Polch} in the matrix base. Rigorous estimates on the
propagator required by the Grosse-Wulkenhaar analysis and a more
explicit multiscale analysis were provided in
\cite{Rivasseau2005bh}. Then Gurau et al. gave another proof that
the non-commutative $\varphi^{\star 4}_4$ is renormalizable, also
with a multiscale analysis but completely in position space
\cite{xphi4-05}. The corresponding parametric representation of the
model was also built in \cite{gurauhypersyman}. Recently the model
has been shown to have no Landau ghost, so that it is actually
better behaved than its commutative counterpart
\cite{GrWu04-2,DisertoriRivasseau2006,beta2-06} and can presumably
be built non perturbatively.

Apart from the $\varphi^{\star 4}_4$ theory, many other theories in
non-commutative space have now also been proved to be renormalizable
to all orders, such as the Gross-Neveu model in 2 dimensional Moyal
plane \cite{RenNCGN05}, the LSZ model \cite{Langmann:2003if} and the
$\varphi^{\star 3}$ theory in various dimensional space
\cite{Grosse2005ig,Grosse2006qv,Grosse2006er}. For an updated
review, see \cite{rivfvt,riv07}.

In this paper we prove that the orientable non-commutative complex
$\varphi^{\star 6}$  field theory, $(\bar\varphi\star\varphi)^3$ for
short, in $2 +1 $ dimensional space, with two dimensions equipped
with non-commutative Moyal product and the third one which commutes
with the two others, is renormalisable to all orders of perturbation
theory. In the first section we derive the propagator and establish
the $x$-space power counting of the theory. In the second section we
prove that the divergent subgraphs can be renormalized by
counterterms of the form of the initial Lagrangian. Our proof, based
solely on $x$ space with multiscale analysis, follows closely the
strategy of \cite{xphi4-05}. For technical reasons, we restrict
ourselves here to the simpler orientable case, but we plan to study
the nonorientable case or real scalar $\varphi^6_3$ model as well.

We are motivated by the fact that the quantum Hall effect at finite
temperature should also be described by a 2+1 dimensional field
theory with two anticommuting space and one commuting imaginary time
coordinates \cite{Suss,Poly,HellRaam,rivfvt}. Our model is therefore
a first step towards understanding how to renormalize such theories.
We plan to compute in a future publication the renormalization group
flow of this model, which involves three parameters $\lambda$, $g$
and $\Omega$, instead of two in the $\varphi^{4}_4$ case.

\section{Power Counting in $x$-Space}
\setcounter{equation}{0}
\subsection{Model, Notations}

The simplest orientable non-commutative complex $\varphi^{\star
6}_3$ theory is defined on ${\mathbb R}^3$ equipped with the
associative and non-commutative Moyal product
\begin{align}
  (a\star b)(x) &= \int \frac{d^2k}{(2\pi)^2} \int d^2 y \; a(x{+}\tfrac{1}{2}
  \theta {\cdot} k)\, b(x{+}y)\, \mathrm{e}^{\mathrm{i} k \cdot y}\;
\label{starprod}
\end{align}

The action functional is
\begin{eqnarray}\label{action}
 S[\varphi] &=& \int \,d^2 x\ d x^0 \Big( \partial_\mu \bar{\varphi} \star
 \partial^\mu \varphi + \partial_0 \bar{\varphi}\star \partial^0 \varphi + \Omega^2 (\tilde{x}_\mu \bar{\varphi} )
 \star (\tilde{x}^\mu \varphi ) +  \mu_0^2 \,\bar{\varphi}
 \star \varphi \nonumber\\
 &+& \frac{\lambda}{2} \bar{\varphi} \star \varphi \star \bar{\varphi} \star
 \varphi + \frac{g}{3} \bar{\varphi} \star \varphi \star \bar{\varphi} \star
 \varphi \star \bar{\varphi} \star \varphi \Big)(x,x^{0})\;
\end{eqnarray}
where $\tilde{x}_\mu=2(\theta^{-1})_{\mu\nu} x^\nu$, and
$x=(x^\mu),\mu=(1,2) $ are the non-commutative variables   and
$x^{0}$ is the commutative variable, that
is\,$$[x^{\mu},x^{\nu}]=i\theta^{\mu \nu} ,\, [x ^{0}, x^{\mu}]=
0.$$
Here $\theta^{\mu \nu}$ is a constant matrix and the Euclidean
metric is used.

\begin{lemma}\label{propagator}
The kernel of the propagator in our $(\bar\varphi\star\varphi)^3$
model is
\begin{equation}\label{propagatoreq}
C(x,x')=\frac{{\Omega}{(2t)^{-\frac{1}{2}}}}{{\sqrt{2\pi}}^3
\sh(2\Omega t) } e^{-\frac{\Omega\coth(2\Omega t)}{2}(x^2+x'^2)+
\frac{\Omega}{\sh(2\Omega t)}x \cdot x'-\frac{(x_{0}-x'_{0})^2}{4t}-
\mu_0^2 t } ,
\end{equation}
with\, $ x^2=x_1^2+x_2^2 , \,{x'}^2={x'}_1^2+{x'}_2^2 ,\,x\cdot
x'={x_1}{x'_1}+{x_2}{x'_2}$.
\end{lemma}

\noindent{\bf Proof} The propagator of interest is expressed via the
Schwinger parameter trick as:
\begin{equation}
  H^{-1}=\int_0^{\infty}\,dt e^{-tH} .
\end{equation}

Let H be
\begin{equation}
H=-\partial_1^2-\partial_2^2-\partial_0^2+
    \Omega^2x^2+\mu_0^2
\end{equation}
where $\mu_0$ is the mass of the field.

The integral kernel of the operator $e^{-tH}$ is:
\begin{equation}
    e^{-tH}(x,x')=\frac{\Omega (2t)^{-\frac{1}{2}}}{{\sqrt{2\pi}}^{3}\sh(2\Omega t)}e^{-A} \ ,
\end{equation}
\begin{equation}
   A=\frac{\Omega\ch(2\Omega t)}{2\sh(2\Omega t)}(x^2+x'^2)-
    \frac{\Omega}{\sh(2\Omega t)}x\cdot
    x'+\frac{(x_0-x'_0)^2}{4t}+\mu_0^2t  \ .
\end{equation}

At first we note that the kernel is correctly normalised: as $\Omega
\rightarrow 0$, we have
\begin{equation}
  e^{-tH}(x,x')\rightarrow\frac{1}{(4\pi t)^{\frac{3}{2}}}e^{-\frac{|x-x'|^2+|x_0-x'_0|^2}{4t}},
\end{equation}
which is the normalised heat kernel. Then we must check the equation
\begin{equation}\label{diffbasic}
  \frac{d}{dt}e^{-tH}+He^{-tH}=0 .
\end{equation}

In fact
\begin{eqnarray}
  \frac{d}{dt}e^{-tH}&=&\frac{\Omega e^{-A} (2t)^{-\frac{1}{2}}}{{\sqrt{2\pi}}^{3}\sh(2\Omega t)}
  \Big{\{}-2\Omega\coth(2\Omega t) +
  \frac{\Omega^2}{\sh^2(2\Omega t)}(x^2+x'^2) \nonumber\\
  &-&{\frac{1}{2}t^{-1}}- \frac{2\Omega^2\ch(2\Omega t)}
  {\sh^2(2\Omega t)}x\cdot x'+\frac{(x_0-x'_0)^2}{4 t^2}-\mu_0^2
  \Big{\}} .
\end{eqnarray}

Moreover
\begin{eqnarray}
\label{equ2.11}
  (-\partial^2_1-\partial^2_2)e^{-tH}
  &=& \frac{\Omega e^{-A} (2t)^{-\frac{1}{2}}}{{\sqrt{2\pi}}^{3}\sh(2\Omega t)}\Big{\{}
  2\Omega\coth(2\Omega t) -\frac{\Omega^2}{\sh^2(2\Omega t)} (x^2+x'^2) \nonumber\\
  &+& \frac{2\Omega^2
  \coth(2\Omega t)}{\sh\Omega t}x\cdot x'
  -\Omega^2x^2
  \Big{\}} , \\
  -\partial^2_0 e^{-tH} &=& \frac{\Omega e^{-A}
(2t)^{-\frac{1}{2}}}{{\sqrt{2\pi}}^{3}\sh(2\Omega t)}\Big{\{}
\frac{1}{2t}-\frac{(x_0-x'_0)^2}{4{t^2}}  \Big{\}} .
\end{eqnarray}

It is now straightforward to verify the differential equation
(\ref{diffbasic}), which proves the lemma.

Now let's consider the interaction vertices.The non-commutative
complex $\varphi^6_3$ model may a priori exhibit both orientable
vertices:
\begin{eqnarray}\label{orientable}
V_{o}&=&\frac{1}{2}\bar{\varphi} \star \varphi \star \bar{\varphi}
\star \varphi(x)
%\nonumber \\&
+\frac{1}{3}\bar{\varphi} \star \varphi \star \bar{\varphi} \star
 \varphi \star \bar{\varphi} \star \varphi(x)
\end{eqnarray}
and non-orientable vertices:
\begin{eqnarray}\label{non-orientable}
V_{no}&=&\frac{1}{2}\bar{\varphi} \star \bar\varphi \star\varphi
\star
 \varphi(x)
 %\nonumber\\&
+\frac{1}{3}\bar{\varphi} \star \bar\varphi \star \bar{\varphi} \star
\varphi \star \varphi \star \varphi(x)
\nonumber\\&
+&\frac{1}{3}\bar{\varphi} \star \bar\varphi \star \varphi \star\bar{\varphi}
 \star\varphi \star \varphi (x)
%\nonumber\\&
 +\frac{1}{3}\bar{\varphi} \star \bar\varphi \star \varphi \star
 \varphi \star \bar\varphi \star \varphi(x) .
\end{eqnarray}
In this paper we limit ourselves to the case of action (\ref{action}), hence
to orientable vertices.

In the two dimensional non-commutative space
the interaction vertices (orientable or not) can be
written as \cite{Filk:1996dm,RenNCGN05}:
\begin{eqnarray}\label{vertexphi4o}
V(x_1, x_2, x_3, x_4) &=& \delta(x_1 -x_2+x_3-x_4)e^{i \sum_{1 \le
i<j \le 4}(-1)^{i+j+1}x_i \theta^{-1}  x_j}
\end{eqnarray}
for four point vertices and
\begin{eqnarray}\label{vertex6o}
 V(x_1, x_2, x_3, x_4, x_5, x_6) &=& \delta(x_1
 -x_2+x_3-x_4+x_5-x_6) \nonumber \\
 && e^{i
 \sum_{1 \le i<j \le 6}(-1)^{i+j+1}x_i \theta^{-1}x_j}
\end{eqnarray}
for six point vertices. Here we note  $x \theta^{-1}  y  \equiv
\frac{2}{\theta} (x_1 y_2 -  x_2  y_1 )$. These vertices are completed to make them local
in the commutative $t$ coordinate, that is we have to multiply
them by $\delta(x^0_1-x^0_2)\delta(x^0_1-x^0_3)\delta(x^0_1-x^0_4)$
or $\delta(x^0_1-x^0_2)\delta(x^0_1-x^0_3)\delta(x^0_1-x^0_4)\delta(x^0_1-x^0_5)\delta(x^0_1-x^0_6)$ respectively.

The main result of this
paper is a proof in configuration space of
\begin{theorem}[BPHZ Theorem for non-commutative $(\bar\varphi\star\varphi)^3$ ]
\label{BPHZ1} The theory defined by the action (\ref{action}) is
renormalizable to all orders of perturbation theory.
\end{theorem}

Let $G$ be an arbitrary connected graph. The amplitude associated
with this graph is (with selfexplaining notations):
\begin{eqnarray}\label{amplitude}
A_G&=&\int \prod_{v \in V_4,i=1,...4} dx_{v,i} dx^{0}_{v}
\prod_{v \in V_6,i=1,...6} dx_{v,i} dx^{0}_{v}\prod_l dt_l     \\
&& \prod_{V_4} {[} \delta(x_{v,1}-x_{v,2}+x_{v,3}-x_{v,4}) e^{\imath
\sum_{i<j}(-1)^{i+j+1}x_{v,i}\theta^{-1} x_{v,j}} {]}\nonumber \\
&& \prod_{V_6} {[}
\delta(x_{v,1}-x_{v,2}+x_{v,3}-x_{v,4}+x_{v,5}-x_{v,6}) e^{\imath
\sum_{i<j}(-1)^{i+j+1}x_{v,i}\theta^{-1} x_{v,j}}{]}
\nonumber  \\
&& \hskip-1cm\prod_l \frac{{\Omega}{(2t_l)^{-\frac{1}{2}}}}{{\sqrt{2\pi}}^3
\sh(2\Omega t_l) } e^{-\frac{\Omega\coth(2\Omega
t_l)}{2}(x_{v,i(l)}^2+x_{v',i'(l)}'^2)+ \frac{\Omega}{\sh(2\Omega
t_l)}x_{v,i(l)} \cdot x'_{v',i'(l)}-\frac{(x_{0,l}-x'_{0,l})^2}{4t_l}-
\mu_0^2 t_l }\;  . \nonumber
\end{eqnarray}

For each line $l$ of the graph joining positions $x_{v,i(l)}$ and
$x_{v',i'(l)}$, we choose an orientation (see next section) and we
define the ``short"
variable $u_l=x_{v,i(l)}-x_{v',i'(l)}$,\\
$u^{0}_l=x^0_{v(l)}-x^0_{v'(l)}$\,and the ``long" variable
$v_l=x_{v,i(l)}+x_{v',i'(l)}$ just as the work of Gurau et al.
\cite{xphi4-05}. With these notations, defining $2\Omega
t_l=\alpha_l$, the propagators in our graph can be written as:
\begin{equation}\label{tanhyp}
\int\prod_l \frac{\sqrt{\Omega}\alpha_l^{-\frac{1}{2}}
d\alpha_l}{2{\sqrt{2\pi}}^3\sinh(\alpha_l)}
e^{-\frac{\Omega}{4}\coth(\frac{\alpha_l}{2})u_l^2-
\frac{\Omega}{4}\tanh(\frac{\alpha_l}{2})v_l^2  -
\frac{\mu_0^2}{2\Omega}
\alpha_l-\frac{\Omega}{2\alpha_l}{u_l^{0}}^2}\; .
\end{equation}

\subsection{Orientation and Position Routing}

\begin{figure}[htbp]
\centering
\includegraphics[scale=1]{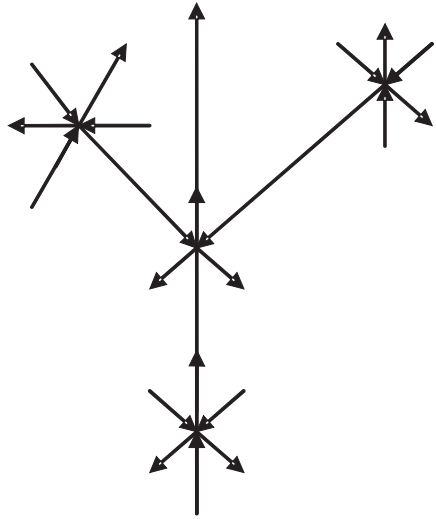}
\caption{Orientation of a tree} \label{fig:treeorient}
\end{figure}

We solve the $\delta$ function at every vertex by a ``position
routing", following the strategy and notations of \cite{xphi4-05}.
The position routing is similar to the ``momentum routing" of the
commutative case, but we have to take care of the cyclic invariance
of the vertex. Consider a connected graph $G$. We choose a rooted
spanning tree in $G$, then we start from an arbitrary orientation of
a first field at the root and inductively climbing into the tree, at
each vertex we follow the cyclic order to alternate entering and
exiting lines. This is pictured in Figure \ref{fig:treeorient}.

Let $n=n_6+n_4$ be the number of vertices of the graph, with $n_6$
of the $\varphi^6$ and $n_4$ of the $\varphi^4$ type, $N$ the number
of its external fields, and $L$ the number of internal lines of $G$.
We have $L=3n_6+2n_4-N/2$.

Every line of the spanning tree by definition has one end exiting a
vertex and one end entering another. This may not be true for the
loop lines, which join two ``loop fields". Among these, some exit
one vertex and enter another; they are called well-oriented. But
others may enter or exit at both ends. These loop lines are
subsequently referred to as ``clashing lines" \cite{xphi4-05}. If
there are no clashing lines, the graph is called orientable. This is
exactly the case in this paper, because $\varphi$ variables can
contract only to $\bar \varphi$ ones. Choosing the $\varphi$
variables as entering and the $\bar \varphi$ as exiting, the form of
the vertices in (\ref{action}) ensure alternance of entering and
exiting lines.

We also define the set of ``branches" associated to the rooted
tree $T$. There are $n-1$ such branches $b(l)$, one for each of the
$n-1$ lines $l$ of the tree. The full tree itself is
called the root branch and noted $b_0$. Each branch is made of
the subgraph $G_b$ containing all the vertices ``above $l$" in $T$,
plus the tree lines and loop lines joining these vertices. It has
also ``external fields" which are the true external fields hooked to
$G_b$, plus the loop fields in $G_b$ for the loops with one end (or
``field") inside and one end outside $G_b$, plus the upper end of
the tree line $l$ itself to which $b$ is associated. We call $X_b$
the set of all external fields $f$ of $b$.

We can now describe the position routing associated to $T$. Here we
will not limit ourselves to orientable graphs but will deal with the
non-orientable graphs as well. There are $n$ $\delta$ functions in
(\ref{amplitude}), hence $n$ linear equations for the $6n_6+4n_4$
positions, one for each vertex. The position routing associated to
the tree $T$ solves this system by passing to another equivalent
system of $n$ linear equations, one for each branch of the tree.
This equivalent system is obtained by summing the arguments of the
$\delta$ functions of the vertices in each branch. To do this we
firstly fix a particular branch $G_b$, with its subtree $T_b$. In
the branch sum we find a sum over all the $u_l$ short parameters of
the lines $l$ in $T_b$ and no $v_l$ long parameters since $l$ both
enters and exits the branch. This is also true for the set $L_b$ of
well-oriented loops lines with both fields in the branch. For the
set $L_{b,+}$ of clashing loops lines with both fields entering the
branch, the short variable disappears and the long variable remains;
the same is true but with a minus sign for the set $L_{b,-}$ of
clashing loops lines with both fields exiting the branch. Finally we
find the sum of positions of all external fields for the branch
(with the signs according to entrance or exit). Obviously the
Jacobian of this transformation is 1, so we simply get another
equivalent set of $n$ $\delta$ functions, one for each branch.

\begin{figure}[htbp]
  \centering
  \includegraphics[scale=0.8]{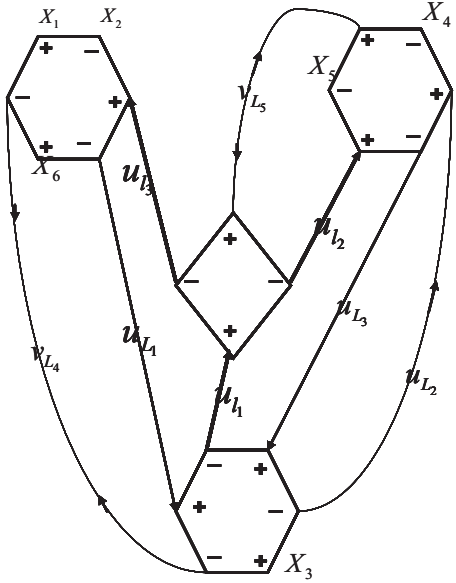}
  \caption{A branch}
  \label{fig:exbranch}
\end{figure}

For instance in the particular case of Figure \ref{fig:exbranch},
the delta function is
\begin{equation}
\delta( u_{l_{1}}+u_{l_{2}}+u_{l_{3}}+u_{L_{1}}+u_{L_{2}}+u_{L_{3}}
 -v_{L_{4}}+v_{L_{5}}+X_{1}-X_{2}+X_{3}-X_{4}-X_{5}+X_{6}) .
\end{equation}

For an orientable graph, the position routing is summarised  by:

\begin{lemma}[Position Routing]
We have, calling $I_G$ the remaining integrand in (\ref{amplitude}):
\begin{eqnarray}
A_G&=&\int \prod_{v_4} {[} \delta(x_{v,1}-x_{v,2}+x_{v,3}-x_{v,4})
{]} \\
&& \prod_{v_6}{[}\delta(x_{v,1}-x_{v,2}+x_{v,3}-x_{v,4}+x_{v,5}-x_{v,6}){]}I_G(\{x_{v,i},x^{0}_{v,i}\})\nonumber\\
&=& \int \prod_{b}\delta \left(\sum_{l\in T_b \cup L_b }u_{l}
 + \sum_{f\in X_b}\epsilon(f)
x_f\right)I_G(\{x_{v,i},x^{0}_{v,i} \})\nonumber
\end{eqnarray}
where $\epsilon(f)$ is $\pm 1$ depending on whether the field $f$
enters or exits the branch.
\end{lemma}

Using the above equations one can at least solve all the long tree
variables $v_l$ in terms of external variables, short variables and
long loop variables, using the $n-1$ non-root branches. There
remains then the root branch $\delta$ function. If $G_b$ is
orientable, this $\delta$ function of branch $b_0$ contains only
short and external variables. Here we shouldn't forget that each
external variable can be written as linear combination of short
variable and long variable. If $G_b$ is non-orientable one can solve
for an additional ``clashing" long loop variable. We can summarise
these observations in the following lemma just like that in
\cite{xphi4-05}:

\begin{lemma}\label{lemmarouting}
The position routing solves any long tree variable $v_l$ as a
function of:
\begin{itemize}
\item the short tree variable $u_l$ of the line $l$ itself,
\item the short tree and loop variables with both ends in $G_{b(l)}$,
\item the short and long variables of the loop lines
with one end inside $G_{b(l)}$ and the other outside,
\item the true external variables $x$ hooked to  $G_{b(l)}$.
\end{itemize}
\end{lemma}

In the orientable case the root branch $\delta$ function contains
only short tree variables, short loop variables and external
variables but no long variables, hence gives a linear relation among
the short variables and external positions. In the non-orientable
case it gives a linear relation between the long variables $w$ of
all the clashing loops in the graph some short variables $u$'s and
all the external positions.

From now on, each time we use this lemma to solve the long tree
variables $v_l$ in terms of the other variables, we shall call $w_l$
rather than $v_l$ the remaining $2n_6+n_4+1 - N/2$ independent long
loop variables. Hence looking at the long variables names the reader
can check whether Lemma \ref{lemmarouting} has been used or not.
\label{filkreduc1}

\subsection{Multiscale Analysis and Crude Power Counting}

In this section we follow the standard procedure of multiscale
analysis \cite{Riv1}. First the parametric integral for the
propagator is sliced in the usual way :
\begin{equation}
C(u,u^{0},v)=C^0(u,u^{0},v)+\sum_{i=1}^{\infty}C^i(u,u^{0},v),
\end{equation}
with
\begin{equation}
C^{0}(u,u^{0},v)=\int_{1}^{\infty}\frac{\sqrt{\Omega}\alpha^{-\frac{1}{2}}
d\alpha}{2{\sqrt{2\pi}}^3\sinh(\alpha)}
e^{-\frac{\Omega}{4}\coth(\frac{\alpha}{2})u^2-
\frac{\Omega}{4}\tanh(\frac{\alpha}{2})v^2  -
\frac{\mu_0^2}{2\Omega} \alpha-\frac{\Omega}{2\alpha_l}{(u^{0})}^2}
\end{equation}
and
\begin{equation}
C^i(u,u^{0},v)=\int_{M^{-2i}}^{M^{-2(i-1)}}\frac{\sqrt{\Omega}\alpha^{-\frac{1}{2}}
d\alpha}{2{\sqrt{2\pi}}^3\sinh(\alpha)}
e^{-\frac{\Omega}{4}\coth(\frac{\alpha}{2})u^2-
\frac{\Omega}{4}\tanh(\frac{\alpha}{2})v^2  -
\frac{\mu_0^2}{2\Omega} \alpha-\frac{\Omega}{2\alpha_l}{(u^{0})}^2} .
\end{equation}

We have an associated decomposition of any amplitude of the theory
as
\begin{equation}
A_{G}=\sum_{\mu}A^{\mu}_{G} .
\end{equation}

\begin{lemma} For some constants $K$ (large) and $c$ (small):
\begin{equation}\label{eq:propbound-phi4}
C^i (u,v) \les K M^{i}e^{-c \big[ M^{i}\Vert u \Vert + M^{i}\Vert u^{0}
\Vert + M^{-i}\Vert v\Vert \big] }
\end{equation}
(which a posteriori justifies the terminology of ``long" and
``short" variables).
\end{lemma}
We can use the second order approximation of the hyperbolic
functions near the origin to prove this lemma.

Taking absolute values, hence neglecting all oscillations, leads to
the following crude bound:
\begin{equation}\label{assignsum}
\vert A_G \vert \les \sum_{\mu}\int\prod_l du_ldu^{0}_ldv_l
C^{i_l}(u_l,u^0_l,v_l) \prod_v \delta_v \ ,
\end{equation}
where $\mu$ is the standard assignment of an integer index $i_l$ to
each propagator of each internal line $l$ of the graph $G$, which
represents its ``scale". We will consider only amputated graphs.
Therefore we have only external vertices of the graph; in the
renormalization group spirit, the convenient convention is to assign
all external indices of these external fields to a  fictitious $-1$
``background" scale.

To any assignment $\mu$ and scale $i$ are associated the standard
connected components $G_k^i$, $k=1,... ,k(i)$ of the subgraph $G^i$
made of all lines with scales $j\ges i$. These components are
partially ordered according to their inclusion relations and the
(abstract) tree describing these inclusion relations is called the
Gallavotti-Nicol\`o tree \cite{GaN,xphi4-05}; its nodes are the
$G_k^i$'s and its root is the complete graph $G$.

More precisely for an arbitrary subgraph $g$ one defines:
\begin{equation}
 i_g(\mu)=\inf_{l\in g}i_l(\mu) \quad , \quad
 e_g(\mu)=\sup_{l \mathrm{~external~line~of~} g}i_l(\mu)\ .
\end{equation}
The subgraph $g$ is a $G_k^i$ for a given $\mu$ if and only if
$i_g(\mu)\ges i> e_g(\mu)$. Now we should choose the real tree $T$
compatible with the abstract Gallavotti-Nicol\`o tree to optimise
the bound over spatial integrations, which means that the
restriction $T_k^{i}$ of $T$ to any $G_k^{i}$ must still span
$G_k^{i}$. This is always possible (by a simple induction from
leaves to root). We pick such a compatible tree $T$ and use it both
to orient the graph as in the previous section and to solve the
associated branch system of $\delta$ functions according to Lemma
\ref{lemmarouting}. We obtain:
\begin{eqnarray}\label{bound1}
\vert A_{G,\mu} \vert \les K^n\prod_l M^{i_l}\int\prod_l du_l
du^{0}_l dv_l  e^{-c \big[ M^{i_l}\Vert u_l \Vert + M^{i_l}\Vert u^{0}
\Vert + M^{-i_l} \Vert v_l \Vert \big]} \prod_b\delta_b   \;
\nonumber\\
\les  K^n\prod_l M^{i_l}\int\prod_l du_l du^{0}_l dw_l  e^{-c \big[
M^{i_l}\Vert u_l \Vert + M^{i_l}\Vert u^{0} \Vert + M^{-i_l} \Vert
v_l (u,w,x) \Vert \big]} \delta_{b_0}\; .
\end{eqnarray}

Then we can find that any long variable integrated at scale $i$
costs $KM^{2i}$ .The integration over the non-commutative short
variable at scale $i$ brings $KM^{-2i}$, and the commutative one
brings $KM^{-i}$ (there is no long variable in the commutative
dimension) so the integration over each tree line at scale $i$
brings a total convergent factor $KM^{-3i}$. The variables
``solved" by the $\delta$ functions bring or cost nothing. For an
orientable graph we should solve the $n-1$ long variables $v_l$'s of
the tree propagators in terms of the other variables, because this
is the maximal number of long variables that we can solve, and they
have highest possible indices because $T$ has been chosen compatible
with the Gallavotti-Nicol\`o tree structure. We should study more
carefully the commutative variable which is the $0th$ dimension of
any tree line of $T$. While the model for the non-commutative
variables is non local, it is local for the commutative variables.
So we can't integrate over all the position variables (or the
equivalent line variables) but have to save one, the root (we name
it $x_{\nu0}$). We will use this point when we perform the
renormalisation where the amputed amplitude of any connected
component depends only on one commutative external position
$x_{\nu0}$. This point is also very important for the power counting
of the non-orientable model as it implies the maximal number of
commutative short variable we can integrate over is $n-1$ not $n$.
Finally we still have the last $\delta_{b_0}$ function (equivalent
to the overall momentum conservation in the commutative case). It is
optimal to use it to solve one external variable (if any) in terms
of all the short variables and the external ones. Since external
variables are typically smeared against unit scale test functions,
this leaves power counting invariant.

We now define $S$ the set of long variables to be solved via the
$\delta$ functions hence the set of $n-1$ tree lines as there are
only orientable graphs in our model.

 Gathering all the corresponding
factors together with the propagators prefactors $M^{i}$ leads to
the following bound:
\begin{equation}
\vert A_{G,\mu}\vert  \  \les \  K^n \prod_l M^{i_l}\prod_{l\in
S}M^{-3i_l } \ .
\end{equation}

In the usual way of \cite{Riv1} we write
\begin{equation}
\prod_{l}M^{i_l}=\prod_{l}\prod_{i=1}^{i_l}M= \prod_{i,k}\prod_{l\in
G^i_k}M=\prod_{i,k}M^{l(G^i_k)}
\end{equation}
and
\begin{equation}
\prod_{l\in S}M^{-3i_l}=\prod_{l\in S}\prod_{i=1}^{i_l}M^{-3}=
\prod_{i,k}\prod_{l\in G^i_k\cap S}M^{-3}
\end{equation}
and we must now only count the number of elements in $G^i_k\cap S$.

As remarked above $G^i_k\cap
S=T^i_k$, and the cardinal of $T^i_k$ is $n(G^i_k)-1$.

Using the fact that $2l(G^i_k)-6n_6(G^i_k)-4n_4=-N(G^i_k)$ we can
summarise these results in the following lemma:

\begin{lemma}\label{crudelemma}
The following bound holds for a connected graph of
$(\bar\varphi\star\varphi)^3$ model (with external arguments
integrated against fixed smooth test functions):
\begin{equation}
\vert A_{G,\mu} \vert \les  K^n \prod_{i,k}M^{-\omega(G^i_k)}
\end{equation}
for some (large) constant $K$, with $\omega(G^i_k)=N(G^i_k)/2+n_4-3$
\end{lemma}
This lemma proves the power counting for orientable graphs. But it is
not yet sufficient for a renormalization theorem to all orders of
perturbation. Indeed only planar graphs with a single broken face look like Moyal
products when their internal indices become much higher than their external ones.
So we must prove that the non-planar graphs or
graphs with more than one broken face have
better power counting than what Lemma \ref{crudelemma}
states. Vertices oscillations should be taken into account to prove that,
and this is done in the next section.

\subsection{Improved Power Counting}

Recall that for any non-commutative Feynman graph $G$ we can define
the genus of the graph, called $g$ and the number of faces ``broken
by external legs", called $B$ \cite{GrWu04-3,Rivasseau2005bh}. For a
general graph, we have $g \ges 0$ and $B\ges 1$.

In the previous section we established that
\begin{equation}
\omega (G) \ges  N/2+n_4 -3 \ , \ {\rm if}\   G\  {\rm orientable} .
\end{equation}
The subgraphs with $g=0$ and $B=1$ are called
\emph{planar regular}. We want to prove that they are the only non-vacuum graphs with
$\omega \les 0$.

It is easy to check that planar regular subgraphs are orientable,
but the converse is not true. To prove that {\it orientable
non-planar} subgraphs or {\it orientable planar} subgraphs with
$B\ges 2$ are irrelevant requires to use a bit of the vertices
oscillations to improve Lemma \ref{crudelemma} and get:

\begin{lemma}\label{improvedbound}
For orientable subgraphs with $g\ges 1$ we have
\begin{equation}\label{improvednonplanar}
\omega (G) \ges  N/2 + n_4 + 1 \; .
\end{equation}
For orientable subgraphs with $g = 0$ and $B\ges 2$ we have
\begin{equation}\label{improvedbrokenfaces}
\omega (G) \ges  N/2 +n_4-1\; .
\end{equation}
\end{lemma}
This lemma is sufficient for the purpose of this paper. It implies
directly that graphs which contain only irrelevant subgraphs have
finite amplitudes which are uniformly bounded by $K^n$, using the
standard method of \cite{Riv1} to bound the assignment sum over
$\mu$ in (\ref{assignsum}).

The rest of this subsection is essentially devoted to the proof of
this Lemma. We return before solving $\delta$ functions, hence to
the $v$ variables. We will need only to compute the oscillations
which are quadratic in the long variables $v$'s to prove
(\ref{improvednonplanar}) and the linear oscillations in $v
\theta^{-1} x$ to prove (\ref{improvedbrokenfaces}). Fortunately an
analog problem was solved in momentum space by Filk and
Chepelev-Roiban \cite{Filk:1996dm,Chepelev:2000hm}, and adapted to
position routing by Gurau et al. \cite{xphi4-05}. We just borrow
from the method of \cite{xphi4-05}. As the procedures for our paper
are almost the same as that for $\varphi^4_4$ in \cite{xphi4-05}, we
reproduce the argument as concisely as possible, and we refer to
\cite{xphi4-05,RenNCGN05} for more details. The short variables are
inessential in this subsection, as the integration of them always
bring about convergent terms. But it is convenient to treat on the
same footing the long $v$ and the external $x$ variables, so we
introduce a new global notation $y$ for all these variables. Then
the vertices rewrite as
\begin{equation}
\prod_v\delta(y_1-y_2+y_3-y_4+y_5-y_6+\epsilon^iu_i)e^{\imath
\big(\sum_{i<j}(-1)^{i+j+1}y_i\theta^{-1}y_j + yQu + uRu \big)} \
\end{equation}
for some inessential signs $\epsilon^i$ and some symplectic matrices
$Q$ and $R$. As there are no oscillations for the commutative
coordinates, there are no Filk moves for them. Since the precise
oscillations in the short $u$ variables is not important to this
problem, we will note in the sequel $E_u$ any linear combination of
the $u$ variables. Let's consider the first Filk reduction
\cite{Filk:1996dm}, which contracts tree lines of the graph. It
creates progressively generalised vertices with even number of
fields. At a given induction step and for a tree line joining two
such generalised vertices with respectively $p$ and $q-p+1$ fields
(suppose $p$ is even and $q$ is odd), we assume by induction that
the two vertices are
\begin{eqnarray}\label{firstfilk1}
&& \delta(y_1-y_2+y_3...-y_p+E_u) \delta(y_p-y_{p+1}+...-y_q+E_u)
\\
&& \hskip-1cm e^{  \imath \big(\sum_{1\les i<j\les p}
(-1)^{i+j+1}y_i\theta^{-1}y_j+ \sum_{p\les i<j\les q}
(-1)^{i+j+1}y_i\theta^{-1}y_j+yQu+ uRu   \big) } \  . \nonumber
\end{eqnarray}
Using the second $\delta$ function we see that:
\begin{equation}\label{solveyp}
y_p=y_{p+1}-y_{p+2}+....+y_q-E_u \ .
\end{equation}
Substituting this expression in the first $\delta$ function we get:
\begin{eqnarray}\label{firstfilk2}
  &&\delta(y_1-y_2+...-y_{p+1}+..-y_q+E_u)
  \delta(y_p-y_{p+1}+...-y_q+E_u)
 \\
 && \hskip-1cm e^{\imath\big( \sum_{1\les i<j\les p} (-1)^{i+j+1}y_i\theta^{-1}y_j+
\sum_{p\les i<j \les q} (-1)^{i+j+1}y_i\theta^{-1}y_j+yQu+ uRu
\big)} \  . \nonumber
\end{eqnarray}

The quadratic terms which include $y_p$ in the exponential are
(taking into account that $p$ is an even number):
\begin{equation}
\sum_{i=1}^{p-1}(-1)^{i+1}y_i\theta^{-1}y_p+\sum_{j=p+1}^q
  (-1)^{j+1}y_p\theta^{-1}y_j \
\end{equation}
Using the expression (\ref{solveyp}) for $y_p$ we see that the
second term gives only terms in $yLu$, as $\theta$ is antisymmetry.
The first term yields:
\begin{equation}
\sum_{i=1}^{p-1}\sum_{j=p+1}^q (-1)^{i+1+j+1}y_i\theta^{-1}y_j=
  \sum_{i=1}^{p-1}\sum_{k=p}^{q-1}(-1)^{i+k+1}y_i\theta^{-1}y_j \ ,
\end{equation}
which reconstitutes the crossed terms, and we have recovered the
inductive form of the larger generalised vertex.

\begin{figure}[htbp]
\centering
\includegraphics[scale=1.3]{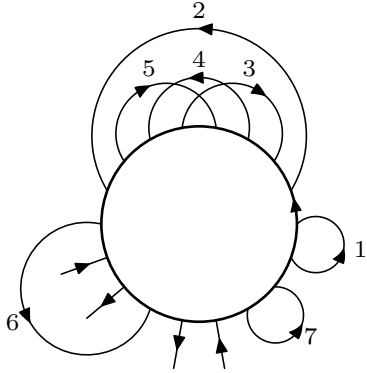}
\caption{A typical rosette} \label{fig:exrosette}
\end{figure}

After each Filk move we will have two more vertices. So by this
procedure we will always treat only even vertices. We finally
rewrite the product of the two vertices as:
\begin{eqnarray}
&&\delta(y_1-y_2+...+y_{p-1}-y_{p+1}+..-y_q+E_u)
\delta(y_p-y_{p-1}+...-y_q+E_u)
\nonumber\\
&&  e^{\imath\big( \sum_{1\les i<j\les
q}(-1)^{i+j+1}y_i\theta^{-1}y_j+yQu+ uRu \big)} ,
\end{eqnarray}
where the exponential is written in terms of the {\it reindexed}
vertex variables. In this way we can contract all lines of a
spanning tree $T$ and reduce $G$ to a single vertex with ``tadpole
loops" called a ``rosette graph" \cite{Chepelev:2000hm}. In this
rosette to keep track of cyclicity is essential so we draw the
rosette as a cycle (which is the border of the former tree) bearing
loops lines on it (see Figure \ref{fig:exrosette}).
Remark that the rosette can also be considered as a big vertex, with
$r=4n_6+2n_4+2$ fields, on which $N$ are external fields with
external variables $x$ and $4n_6+2n_4+2-N$ are loop fields for the
corresponding $2n_6+n_4+1-N/2$ loops. When the graph is orientable,
 the long variables $y_l$ for $l$ in $T$ will
disappear in the rosette. Let us call $z$ the set of remaining long
loop and external variables. Then the rosette vertex factor is
\begin{equation}\label{rosettefactor}
\delta(z_1-z_2+...-z_r+E_u) e^{\imath\big(\sum_{1\les i<j\les
r}(-1)^{i+j+1}z_i\theta^{-1}z_j+zQu+ uRu\big)} \ .
\end{equation}

We can go on performing inductively the first Filk move and the net
effect is simply to rewrite the root branch $\delta$ function and
the combination of all vertices oscillations (using the other
$\delta$ functions) as the new big vertex or rosette factor
(\ref{rosettefactor}).

The second Filk reduction \cite{Filk:1996dm} further simplifies the
rosette factor by erasing the loops of the rosette which do not
cross any other loops or arch over external fields. Putting together
all the terms in the exponential which contain $z_l$ we conclude
exactly as in \cite{Filk:1996dm} that these long $z$ variables
completely disappear from the rosette oscillation factor, which
simplifies as in \cite{Chepelev:2000hm} to
\begin{equation}\label{rosettefactorsimp}
\delta(z_1-z_2+...-z_r+E_u) e^{\imath\lbt z {\cal I}z+zQu+ uRu\rbt}
\;
\end{equation}
where ${\cal I}_{ij}$ is the antisymmetric ``intersection matrix" of
\cite{Chepelev:2000hm} (up to a different sign convention). Here
${\cal I}_{ij}= +1$ if oriented loop line $i$ crosses oriented loop
line $j$  coming from its right, ${\cal I}_{ij} = -1$ if $i$ crosses
$j$  coming from its left, and ${\cal I}_{ij} = 0$ if $i$ and $j$ do
not cross. These formulas are also true for $i$ external line and
$j$ loop line or the converse, provided one extends the external
lines from the rosette circle radially to infinity to see their
crossing with the loops. Finally when $i$ and $j$ are external lines
one should define ${\cal I}_{ij} = (-1)^{p+q+1}$ if $p$ and  $q$ are
the numbering of the lines on the rosette cycle (starting from an
arbitrary origin).

If a node $G^i_{k}$ of the Gallavotti-Nicol\`o tree is orientable
but non-planar ($g \ge 1$), there must therefore exist at least two
intersecting loop lines in the rosette corresponding to this
$G^i_k$, with long variables $w_1$ and $w_2$. Moreover since
$G^i_{k}$ is orientable, none of the long loop variables associated
with these two lines belongs to the set $S$ of long variables
eliminated by the $\delta$ constraints. Therefore, after integrating
the variables in $S$ the basic mechanism to improve the power
counting of a single non planar subgraph is the following:
\begin{eqnarray}\label{gainoscill}
&&\int dw_1dw_2 e^{-M^{-2i_1}w_1^2-M^{-2i_2}w_2^2 -
iw_1\theta^{-1}w_2+w_1 . E_1(x,u)+w_2 E_2(x,u)}
\nonumber\\
&=& \int dw'_1dw'_2 e^{-M^{-2i_1}(w_1')^2 -M^{-2i_2}(w'_2)^2
+iw'_1\theta^{-1}w'_2 + (u,x)Q(u,x)}
\nonumber\\
&=&  K  M^{2i_1} \int dw'_2 e^{- (M^{2i_1}+ M^{-2i_2})(w'_2)^2 }= K
M^{2i_1}\frac{M^{-2i_1}}{1+M^{-2(i_1+i_2)}}\les K .
\end{eqnarray}
In these equations we used for simplicity $M^{-2i}$ instead of the
correct but more complicated factor $(\Omega /4) \tanh (\alpha /2 )$
(see \ref{tanhyp}) (of course this does not change the argument) and
we performed a unitary linear change of variables $w'_1 = w_1 +
\ell_1 (x, u)$, $w'_2 = w_2 + \ell_2 (x, u)$ to compute the
oscillating $w'_1$ integral. The gain in (\ref{gainoscill}) is
$M^{-2i_1-2i_2}$, which is the difference between $O(1)$ and the
normal factor $M^{2i_1+2i_2}$ that would be generated by the
integrals over $w_1$ and $w_2$ if there were not the oscillation
term $iw_1\theta^{-1}w_2$.

So after the integration of the non-commutative part of the two
clashing lines the gain is almost $M^{-4i}$.

This basic argument must then be generalised to each non-planar leaf
in the Gallavotti-Nicol\`o tree. Actually, in any orientable
non-planar `primitive" $G^i_k$ node (i.e. not containing sub
non-planar nodes) we can choose an arbitrary pair of crossing loop
lines which will be integrated as in (\ref{gainoscill}) using this
oscillation. The corresponding improvements are independent.

This leads to an improved amplitude bound:
\begin{equation}
\vert A_{G,\mu} \vert \les K^n \prod_{i,k}M^{-\omega(G^i_k)}\
\end{equation}
where now $\omega(G^i_k)=N(G^i_k)/2+ n_4 + 1$ if $G^i_k$ is
orientable and non planar (i.e. $g \ges 1$). This bound proves
(\ref{improvednonplanar}).

Finally it remains to consider the case of nodes $G^i_k$ which are
planar orientable but with $B \ges 2$. In that case there are no
crossing loops in the rosette but there must be at least one loop
line arching over a non trivial subset of external legs in the
$G^i_k$ rosette (see line 6 in Figure \ref{fig:exrosette}). We have
then a non trivial integration over at least one external variable,
called $x$, of at least one long loop variable called $w$. This
``external" $x$ variable without the oscillation improvement would
be integrated with a test function of scale 1 (if it is a true
external line of scale $1$) or better (if it is a higher long loop
variable)\footnote{Since the loop line arches over a non trivial
(i.e. neither full nor empty) subset of external legs of the
rosette, the variable $x$ cannot be the full combination of external
variables in the ``root" $\delta$ function.}. But we get now
\begin{eqnarray}\label{gainoscillb}
&&\int dx dw e^{-M^{-2i}w^2 - iw\theta^{-1}x  +w.E_1(x',u)}
\nonumber\\
&=&  K  M^{2i} \int dx e^{-M^{+2i} x^2 }= K' \ .
\end{eqnarray}
We find that a factor $M^{2i}$ in the former bound becomes $O(1)$
hence is improved by $M^{-2i}$. So the power counting is
$\omega(G^i_k)=N(G^i_k)/2-1+n_4$ .We find that the two point graphs
with $n_4=0$ and $N(G^{i}_{k})=2$ maybe logarithmically divergent.
They do not appear renormalizable at first sight. But we remark that
{\textit{ in the orientable $(\bar\varphi\star\varphi)^3$ model}}
there will never be such subgraphs with $N(G^{i}_{k})=2$ and $B=2$.
This is the reason we limit ourselves to this case\footnote{We thank
our referee for correcting an earlier version of this paper, which
lead us to this important point.}. Then all graphs with $B\ges 2$
are also safe. The only divergent graphs which need renormalization
are the planar regular graphs.

\section{Renormalization}
\setcounter{equation}{0}

In this section we need to consider only divergent subgraphs, namely
the planar two point, four point and six point subgraphs with a
single external face ($g=0$, $B=1$, $ N=2 ,\,4 ,\, 6 \,$  for
$\,n_4=0 $\, ,$N=2 ,\, 4 \,$ for $\,n_4=1 $, and  \,$ N=2 $ for
$\,n_4=2 $). We shall prove that they can be renormalized by
appropriate counterterms of the form of the initial Lagrangian. We
would like to remark that for any graph, contrary to the
non-commutative variables, the commutative variables of the external
points are local. So there is only one integral over the commutative
variable for each vertex.

\subsection{Renormalization of the Six-point Function}\label{ren6pt}
Consider a 6 point subgraph which needs to be renormalized, hence is
a node of the Gallavotti-Nicol\`o tree. This means that there is
$(i,k)$ such that $N(G^{i}_{k})=6$. The six external positions of
the amputated graph are labelled $x_{1},x_{2},x_{3},x_{4},x_{5}$ and
$x_{6}$. We also define $Q$, $R$ and $S$ as three skew-symmetric
matrices of respective sizes $6\times l(G^{i}_{k})$,
$l(G^{i}_{k})\times l(G^{i}_{k})$ and $2[n_6(G^{i}_{k})-1]\times
l(G^{i}_{k})$, where we recall that $2(n(G)_6-1)$ is the number of
loops of a 6 point graph with $n_6$ vertices. The amplitude
associated to the connected component $G^{i}_{k}$ is then
\begin{eqnarray}
A(G^{i}_{k})(x_{1},x_{2},x_{3},x_{4},x_{5},x_{6},x^{0}_{\nu})
&=&\int\prod_{\ell\in T^{i}_{k}} du_{\ell} du^{0}_{\ell} C_{\ell}(x,
u, u^{0}, w)\nonumber\\
\nonumber\\
&&\hskip-6.5cm\prod_{l \in G^{i}_{k},\,l\not\in T} du_{l} du^{0}_l d
w_{l}
C_{l}(u_l,u^{0}_l,w_l)\delta\Big(x_{1}-x_{2}+x_{3}-x_{4}+x_{5}-x_{6}+\sum_{l\in
G^{i}_{k}} u_l\Big)\nonumber\\
&&\hskip-4cm\times e^{\imath \lbt
\sum_{p<q}(-1)^{p+q+1}x_{p}\theta^{-1} x_{q}+XQU+URU+USW \rbt } .
\label{eq:4pt-ini1}
\end{eqnarray}
Here the variable $x^{0}_{\nu}$ is the root commutative variable as
discussed in section ($2.3$) and we will write it as $x^0$
hereafter. The exact form of the factor
$$\sum_{p<q}
(-1)^{p+q+1}x_{p}\theta^{-1} x_{q}$$
is not essential for this paper
and was discussed exhaustively in \cite{xphi4-05,RenNCGN05}.
The important fact is that there are no
quadratic oscillations in $X$ times $W$ (because $B=1$) nor in $W$
times $W$ (because $g=0$). $C_{l}$ is the propagator of the line
$l$. For loop lines $C_{l}$ is expressed in terms of $u_l$ and $w_l$
by formula (\ref{tanhyp}), (with $v$ replaced by our notation $w$
for long variables of loop lines). But for tree lines $\ell \in
T^i_k$ recall that the solution of the system of branch $\delta$
functions for $T$ has reexpressed the corresponding long variables
$v_\ell$ in terms of the short variables $u$, and the external and
long loop variables of the branch graph $G_\ell$ which lies ``over"
$\ell$ in the rooted tree $T$. This is the essential content of
subsection \ref{filkreduc1}. More precisely consider a line $\ell
\in T^i_k$ with scale $i(\ell)\ges i$; we can write
\begin{equation}\label{exactvvalue}
v_\ell =  X_\ell + W_\ell + U_\ell
\end{equation}
where
\begin{equation}\label{xvalue}
X_\ell = \sum_{e\in E(\ell)}  \epsilon_{\ell,e} x_{e}
\end{equation}
is a linear combination on the set of external variables of the
branch graph $G_\ell$ with the correct alternating signs
$\epsilon_{\ell,e}$,
\begin{equation}\label{wvalue}
W_\ell = \sum_{l \in \calL (\ell)} \epsilon_{\ell,l} w_{l}
\end{equation}
is a linear combination over the set $\calL (\ell)$ of long loop
variables for the external lines of $G_\ell$ (and
$\epsilon_{\ell,l}$ are other signs), and
\begin{equation}\label{uvalue}
U_\ell = \sum_{l' \in S (\ell)} \epsilon_{\ell, l'} u_{l'}
\end{equation}
is a linear combination over a set $S_\ell$ of short variables that
we do not need to know explicitly. The tree propagator for line
$\ell$ then is
\begin{equation}\label{propatre}
C_{\ell}(u_\ell, X_\ell, U_\ell, W_\ell,u^{0}_\ell) =
\int_{M^{-2i(\ell)}}^{M^{-2(i(\ell)-1)}}\hskip-.4cm
\frac{\sqrt{\Omega}\alpha_l^{-\frac{1}{2}} d\alpha_l
e^{-\frac{\Omega}{4} \{ \coth(\frac{\alpha_\ell}{2})u_l^2  +
\tanh(\frac{\alpha_\ell}{2}) [X_\ell + W_\ell + U_\ell ]^2
\}-\frac{u_{0}^2}{2\alpha}\Omega }}{2{\sqrt{2\pi}}^3\sinh(\alpha_l)} .
\end{equation}
To renormalize, let us call $e= \max e_p$, $p=1,...,6 $ the highest
external index of the subgraph $G^{i}_{k}$.  We have $e<i$ since
$G^{i}_{k}$ is a node of the Gallavotti-Nicol\`o tree. We evaluate
$A(G^{i}_{k})$ on external fields\footnote{For the external index to
be exactly $e$ the external smearing factor should be in fact
$\prod_{p} \varphi^{\les e}(x_p) - \prod_{p} \varphi^{\les
e-1}(x_p)$ but this subtlety is inessential.} $\bar\varphi^{\les
e}(x_p,x^{0})$ and $\varphi^{\les e}(x_p,x^{0})$ as:
\begin{eqnarray}
A(G^{i}_{k})&=&\int\prod_{p=1}^{6}dx_{p}dx^{0}\bar{\varphi}^{\les
e}(x_{1},x^{0})\varphi^{\les e}(x_{2},x^{0})\bar{\varphi}^{\les
e}(x_{3},x^{0})
 \varphi^{\les
e}(x_{4},x^{0})\nonumber\\
&\times&\bar{\varphi}^{\les e}(x_{5},x^{0})
\varphi^{\les e}(x_{6},x^{0})A(G^{i}_{k})(x_{1},x_{2},x_{3},x_{4},x_{5},x_{6},x^{0})\nonumber\\
&=&\int\prod_{p=1}^{6}dx_{p }dx^{0} \varphi^{\le e}(x_{p},x^{0})\
e^{\imath\text{Ext}} \prod_{\ell\in T^{i}_{k} } du_{\ell}
du^{0}_{\ell} C_{\ell}(u_\ell,u^{0}_\ell, tX_\ell U_\ell, W_\ell)\\
&\times&\prod_{l \in G^{i}_{k} \, \ l \not \in T}  du_{l} du^{0}_{l}
d w_{l} C_{l}(u_l,u^{0}_l, w_l) \delta\Big(\Delta+t\sum_{l\in
G^{i}_{k}}u_{l}\Big) e^{\imath tXQU+\imath URU+\imath USW}\
\Bigg|_{t=1}\  . \nonumber
\end{eqnarray}
with $\Delta = x_{1}-x_{2}+x_{3}-x_{4}+x_{5}-x_{6}$ and
$\text{Ext}=\sum_{p<q=1}^{6}(-1)^{p+q+1}x_{p}\theta^{-1} x_{q}$.
This formula is designed so that at $t=0$ all dependence on the
external variables $x$ factorizes out of the $u,w$ integral in the
desired vertex form for renormalization of the $\bar\varphi \star
\varphi \star \bar \varphi \star \bar \varphi \star \varphi \star
\varphi$ interaction in the action (\ref{action}). We now perform a
Taylor expansion to first order with respect to the $t$ variable and
prove that the remainder term is irrelevant. Let
$\mathfrak{U}=\sum_{l\in G^{i}_{k}}u_{l}$, and
\begin{eqnarray}\label{rax}
{\mathfrak R}(t) &=&   -\sum_{\ell\in T^{i}_{k}
}\frac{\Omega}{4}\tanh(\frac{\alpha_\ell}{2}) \Bigg\{ t^2 X_\ell ^2
+ 2t X_\ell \big[ W_\ell + U_\ell \big] \Bigg\}
\nonumber\\
&\equiv& - t^2  {\cal A} X . X  -  2t {\cal A}X . (W + U ) \ .
\end{eqnarray}
where ${\cal A}_\ell =
\frac{\Omega}{4}\tanh(\frac{\alpha_\ell}{2})$, and $X\cdot Y$ means
$\sum_{\ell\in T^{i}_{k} } X_\ell  . Y_\ell $. We have
\begin{eqnarray}
A(G^{i}_{k})&=&\int\prod_{p=1}^{6}dx_{p }dx^{0} \varphi^{\le
e}(x_{p},x^{0}) e^{\imath\text{Ext}} \prod_{\ell\in T^{i}_{k} }
du_{\ell} du^{0}_{\ell} C_{\ell}(u_\ell,u^{0}_\ell, U_\ell, W_\ell)
\nonumber\\
&& \bigg[ \prod_{l \in G^{i}_{k} \, \ l \not \in T}  du_{l} du^{0}_l
d w_{l} C_{l}(u_l, w_l) \bigg] \ e^{\imath URU+\imath USW}
\\
&& \hskip -1cm\Bigg\{ \delta(\Delta) + \int_{0}^{1}dt\bigg[ \mathfrak{U}\cdot
\nabla \delta(\Delta+t\mathfrak{U})
+\delta(\Delta+t\mathfrak{U}) [\imath XQU  + {\mathfrak R}' (t)]
\bigg] e^{\imath tXQU + {\mathfrak R}(t)}  \Bigg\} \ \nonumber
\end{eqnarray}
where $C_{\ell}(u_\ell, U_\ell, W_\ell) $ is given by
(\ref{propatre}) but taken at $X_\ell=0$.

The first term, denoted by $\tau A$, is of the desired form
(\ref{vertex6o}) times a number independent of the external
variables $x$. It is asymptotically constant in the slice index $i$,
hence the sum over $i$ at fixed $e$ is logarithmically divergent:
this is the divergence expected for the six-point function. It
remains only to check that $(1-\tau)A$ converges as $i-e \to
\infty$.  But we have three types of terms in $(1-\tau)A$, each
providing a specific improvement over the regular, log-divergent
power counting of $A$:
\begin{itemize}

\item The term $\mathfrak{U}\cdot \nabla \delta(\Delta+t\mathfrak{U})\,
$. For this term, integrating by parts over external variables, the
$\nabla $ acts on external fields $\varphi^{\les e}$, hence brings
at most $M^{e}$ to the bound, whether the $\mathfrak{U}$ term brings
at least $M^{-i}$.

\item  The term $XQU$. Here $X$ brings at most $M^e$ and $U$ brings at least $M^{-i}$.

\item The term ${\mathfrak R}' (t)$. It decomposes into terms in ${\cal A} X \cdot X$, ${\cal A} X\cdot U$
and ${\cal A} X\cdot W$. Here the ${\cal A}_\ell$ brings at least
$M^{-2 i(\ell)}$, $X$ brings at worst $M^{e}$, $U$ brings at least
$M^{-i}$ and $X_\ell  W_\ell$ brings at worst $M^{e+i(\ell)}$. This
last point is the only subtle one: if $\ell \in T^i_k$, remark that
because $T^i_k$ is a sub-tree within each Gallavotti-Nicol\`o
subnode of $G^i_k$, in particular all parameters $w_{l'}$ for $l'
\in\calL (\ell)$ which appear in $W_\ell$ must have indices lower or
equal to $i(\ell)$ (otherwise they would have been chosen instead of
$\ell$ in $T^i_k$).

\end{itemize}

In conclusion, since $i(\ell) \ges i$, the Taylor remainder term
$(1-\tau)A$ improves the power-counting of the connected component
$G_{k}^{i}$ by a factor at least $M^{-(i-e)}$. This additional
$M^{-(i-e)}$ factor makes $(1-\tau)A(G^{i}_{k})$ convergent and
irrelevant as desired.

\subsection{Renormalization of the Four-point Function}\label{Ren4pt}

Consider a 4 point subgraph which needs to be renormalized, hence is
a node of the Gallavotti-Nicol\`o tree. This means that there is
$(i,k)$ such that $N(G^{i}_{k})=4$. The four external positions of
the amputated graph are labelled $x_{1},x_{2},x_{3}$ and $x_{4}$. We
also define $Q$, $R$ and $S$ as three skew-symmetric matrices of
respective sizes $4\times l(G^{i}_{k})$, $l(G^{i}_{k})\times
l(G^{i}_{k})$ and $[2n_6(G^{i}_{k})+n_4(G^{i}_{k})-1]\times
l(G^{i}_{k})$, where we recall that
$[2n_6(G^{i}_{k})+n_4(G^{i}_{k})-1]$ is the number of loops of a 4
point graph with $n_6+n_4$ vertices. The amplitude associated to the
connected component $G^{i}_{k}$ is then
\begin{eqnarray}
&&A(G^{i}_{k})(x_{1},x_{2},x_{3},x_{4},x^{0})\label{eq:4pt-ini2}\\&=&\int
\prod_{\ell\in T^{i}_{k}}  du_{\ell} du^{0}_{\ell}  C_{\ell}(x, u,
u^{0}, w) \prod_{l \in G^{i}_{k},\,l\not\in T} du_{l} du^{0}_l
dw_{l} C_{l}(u_l,u^{0}_l, w_l)
\nonumber\\
&\times&\delta\Big(x_{1}-x_{2}+x_{3}-x_{4}+\sum_{l\in G^{i}_{k}}
u_l\Big) e^{\imath \lbt \sum_{p<q}(-1)^{p+q+1}x_{p}\theta^{-1}
x_{q}+XQU+URU+USW \rbt }. \nonumber
\end{eqnarray}
The renormalization procedure is almost the same as that for the 6
point function. Let us call $e= \max e_p$, $p=1,...,4 $ the highest
external index of the subgraph $G^{i}_{k}$. We have $e<i$ since
$G^{i}_{k}$ is a node of the Gallavotti-Nicol\`o tree. We evaluate
$A(G^{i}_{k})$ on external fields $\bar\varphi^{\les e}(x_p,x^{0})$
and $\varphi^{\les e}(x_p,x^{0})$ as:
\begin{eqnarray}
A(G^{i}_{k})&=&\int\prod_{p=1}^{4}dx_{p}dx^{0}\bar{\varphi}^{\les
e}(x_{1},x^{0})\varphi^{\les e}(x_{2},x^{0})\bar{\varphi}^{\les
e}(x_{3},x^{0})
 \varphi^{\les
e}(x_{4},x^{0})\nonumber\\
&\times&A(G^{i}_{k})(x_{1},x_{2},x_{3},x_{4},x^{0})\nonumber\\
&=&\int\prod_{p=1}^{4}dx_{p }dx^{0} \bar{\varphi}^{\les
e}(x_{1},x^{0})\varphi^{\les e}(x_{2},x^{0})\bar{\varphi}^{\les
e}(x_{3},x^{0})
 \varphi^{\les
e}(x_{4},x^{0})\ e^{\imath\text{Ext}}\nonumber\\
& & \prod_{\ell\in T^{i}_{k} } du_{\ell}
du^{0}_{\ell} C_{\ell}(u_\ell,u^{0}_\ell, tX_\ell U_\ell, W_\ell)\\
&& \prod_{l \in G^{i}_{k} \, \ l \not \in T}  du_{l} du^{0}_l d
w_{l} C_{l}(u_l,u^{0}_l,w_l) \delta\Big(\Delta+t\sum_{l\in
G^{i}_{k}}u_{l}\Big) e^{\imath tXQU+\imath URU+\imath USW}\
\Bigg|_{t=1}\  \nonumber
\end{eqnarray}
with $\Delta =x_{1}-x_{2}+x_{3}-x_{4}$
and\,$\text{Ext}=\sum_{p<q=1}^{4}(-1)^{p+q+1}x_{p}\theta^{-1}
x_{q}$. Then we have

\begin{eqnarray}
 A(G^{i}_{k})&=&\int\prod_{p=1}^{4}dx_{p }dx^{0} \varphi^{\les
 e}(x_{p},x^{0})\ e^{\imath\text{Ext}} \prod_{\ell\in T^{i}_{k} } du_{\ell}
 du^{0}_{\ell} C_{\ell}(u_\ell,u^{0}_\ell, U_\ell, W_\ell)
 \nonumber\\
 && \bigg[ \prod_{l \in G^{i}_{k} \, \ l \not \in T}  du_{l} du^{0}_l d
 w_{l} C_{l}(u_l,u^{0}_l, w_l) \bigg] \ e^{\imath URU+\imath
 USW}\nonumber\\
 & &\Bigg\{ \delta(\Delta) + \mathfrak{U}^{\mu}\cdot \nabla_{\mu}
 \delta(\Delta) + \big[i X Q U-2AX(W+U)\big]\times\delta(\Delta)\nonumber\\
 \hskip-0.5cm&+&\frac{1}{2} \int_{0}^{1}dt(1-t)\bigg[ (\mathfrak{U}\cdot
 \nabla)^{2}
 \delta(\Delta+t\mathfrak{U})+f''(t)\delta(\Delta+t\mathfrak{U})
 \bigg] e^{\imath tXQU + {\mathfrak R}(t)}  \Bigg\} \nonumber\\
 &&
\end{eqnarray}
where $\mathfrak R$ ,$X$ and $\mathfrak U$ are the same as
(\ref{rax}), and again $C_{\ell}(u_\ell,u^{0}_\ell,U_\ell, W_\ell)$
is given by (\ref{propatre}) but taken at $X_\ell=0$.

The first term, denoted by $\tau A$, is of the desired form
(\ref{vertexphi4o}) times a number independent of the external
variables $x$. It is
 is linearly divergent: this is the divergence
expected for the four-point function. It remains only to check that
$(1-\tau)A$ converges as $i-e \to \infty$.  But we have three types
of terms in $(1-\tau)A$, each providing a specific improvement over
the regular, log-divergent power counting of $A$:

\begin{itemize}

\item The term $\mathfrak{U}\cdot \nabla \delta(\Delta)$
 vanishes due to the parity, as it is odd integral over $u$.

\item the third term (the terms linearly proportional to $U$ and $W$ )on the r.h.s. is zero due to the
parity, as they are also odd integrals over $u$ and $w$.

\item In the remainder terms of tailor expansion, for the term $\mathfrak{U}^{2}\cdot
\nabla^{2} \delta(\Delta+t\mathfrak{U})$ the$\nabla^{2}$ brings
$M^{2e}$ through the integral by parts and the $\mathfrak{U}'^{2}$
brings $M^{-2i}$. So it is convergent.

\item for the last term, $f''(t)=-2A.X.X+(X
QU)^{2}+4[AX(W+U)]^{2}\\-4iXQUAX(W+U) $ and this term is convergent.

\end{itemize}

\subsection{Renormalization of the Two-point Function}\label{Ren2pt}

We consider now the nodes such that $N(G^{i}_{k})=2$. We use the same
notations than in the previous subsection. The two external points
are labelled $x$ and $y$. Using the global $\delta$ function, which
is now $\delta\Big(x-y + {\mathfrak U}\Big)$, we remark that the
external oscillation $e^{\imath x \theta^{-1} y}$ can be absorbed in
a redefinition of the term $e^{\imath tXQU}$, which we do from now
on. The full amplitude is

\begin{eqnarray}\label{2point1}
A(G^{i}_{k}) &=&\int dx dy dx^{0}\bar\varphi^{\les
e}(x,x^{0})\varphi^{\les e}(y,x^{0}) \delta\Big(x-y + {\mathfrak
U}\Big) \prod_{l \in G^{i}_{k} ,\; l \not \in T}  du_{l} du^{0}_l d
w_{l}
\nonumber\\
&&C_{l}(u_l,u^{0}_l,w_l)\prod_{\ell\in T^{i}_{k} }  du_{\ell}
du^{0}_{\ell} C_{\ell}(u_\ell,u^{0}_\ell, X_\ell, U_\ell, W_\ell) \
e^{\imath XQU+\imath URU+\imath USW}\; . \nonumber
\end{eqnarray}

We first perform the Taylor expansion in the position variables of
external fields:

\begin{eqnarray}\label{exte}
& &\bar\varphi^{\les e}(x,x^{0})\varphi^{\les e}(y,x^{0})
\delta\Big(x-y + {\mathfrak U}\Big)\nonumber\\&=&\bar\varphi^{\les
e}(x,x^{0})\varphi^{\les e}(y,x^{0})
\delta\Big(x-y + s\mathfrak U\Big)|_{s=1}\nonumber\\
&=&\bar\varphi^{\les e}(x,x^{0})\varphi^{\les
e}(y,x^{0})\Big[\delta(x-y)+\mathfrak{U}\cdot\nabla\delta(x-y)\nonumber\\
&+& \frac{1}{2}(\mathfrak{U}\cdot\nabla)^{2}\delta(x-y
)+\frac{1}{2}\int_{0}^{1} ds
(1-s)^2(\mathfrak{U}\cdot\nabla)^{3}\delta\Big(x-y + s\mathfrak
U\Big)\Big] .
\end{eqnarray}

We perform then a Taylor expansion in $t$ at order $3$ of the
remaining function
\begin{equation}
  \label{eq:f}
  f(t)=  e^{\imath tXQ U   + {\mathfrak R}(t)}\; ,
\end{equation}
where we recall that ${\mathfrak R}(t)= - [ t^2  {\cal A} X . X + 2t
{\cal A}X . (W + U )]$. We get

\begin{eqnarray}\label{2point2}
A_0 &=&  \int dx dx^{0} \bar\varphi^{\les e}(x,x^{0})\varphi^{\les
e}(x,x^{0}) \, e^{\imath (URU+ USW)}
\nonumber\\
&& \prod_{l \in G^{j}_{k} , \; l \not \in T}  du_{l} du^{0}_l d
w_{l} C_{l}(u_l, w_l)
 \prod_{\ell\in T^{i}_{k}}  du_{\ell} du^{0}_\ell  C_{\ell}(u_\ell,u^{0}_\ell, U_\ell, W_\ell)
\nonumber\\
&& \lbt f(0)+f'(0)+\frac 12f''(0)+\frac
12\int_{0}^{1}dt\,(1-t)^{2}f^{(3)}(t)\rbt\  . \
\end{eqnarray}

In order to evaluate that expression, let $A_{0,0},A_{0,1},A_{0,2}$
be the zeroth, first and second order terms in this Taylor
expansion, and $A_{0,R}$ be the remainder term. First,

\begin{eqnarray}\label{2point2bis}
A_{0,0}&=& \int dx dx^{0} \bar\varphi^{\les e}(x,x^{0})\varphi^{\les
e}(x,x^{0}) \, e^{\imath (URU+ USW)}
\nonumber\\
&& \prod_{l \in G^{j}_{k} , \; l \not \in T}  du_{l} du^{0}_l d
w_{l} C_{l}(u_l, w_l)
 \prod_{\ell\in T^{i}_{k}}  du_{\ell} du^{0}_\ell C_{\ell}(u_\ell,u^{0}_\ell, U_\ell, W_\ell)
\end{eqnarray}
is quadratically divergent and is exactly the expected form for the
mass counterterm. Then
\begin{eqnarray}
A_{0,1}&=& \int dx dx^{0} \bar\varphi^{\les e}(x,x^{0})\varphi^{\les
e}(x,x^{0}) \, e^{\imath (URU+ USW)} \prod_{l \in G^{i}_{k}  ,\; l
\not \in T} du_{l}  d w_{l} C_{l}(u_l, w_l)
\nonumber\\
&& \prod_{\ell\in T^{i}_{k} }  du_{\ell} du^{0}_{\ell}
C_{\ell}(u_\ell,u^{0}_\ell, U_\ell, W_\ell) \bigg(  \imath XQU  +
{\mathfrak R}' (0)   \bigg)
\end{eqnarray}
vanishes identically. Indeed all the terms are odd integrals over
the $u$ and $w$ variables. $A_{0,2}$ is more complicated:
\begin{eqnarray}\label{eqtwo2}
A_{0,2}&=& \int dx dx^{0} \bar\varphi^{\les e}(x,x^{0})\varphi^{\les
e}(x,x^{0}) \, e^{\imath (URU+ USW)}
\prod_{l \in G^{i}_{k}  ,\; l \not \in T}  du_{l}  d w_{l} C_{l}(u_l, w_l)\nonumber\\
&& \prod_{\ell\in T^{i}_{k} }  du_{\ell}   C_{\ell}(u_\ell, U_\ell,
W_\ell)  \ \Bigg( -( XQU)^2
\nonumber\\
&& \hskip -1cm- 4\imath XQU {\cal A}X \cdot (W + U )  -2 {\cal A}X
\cdot X +4  [{\cal A}X \cdot (W + U )] ^2 .
 \Bigg)
\end{eqnarray}

The four terms in $(XQU)^2$, $XQU {\cal A}X \cdot W$, ${\cal A}X
\cdot X$ and  $[{\cal A}X \cdot W ] ^2 $ are logarithmically
divergent and contribute to the renormalization of the  harmonic
frequency term $\Omega$ in (\ref{action}). (The terms in $x^\mu
x^\nu$ with  $\mu \ne \nu$ do not survive by parity and the terms in
$(x^\mu )^2$  have obviously the same coefficient.) The other terms
in $XQU {\cal A}X \cdot U$,
 $({\cal A}X \cdot U)({\cal A}X \cdot W)$ and  $[{\cal A}X \cdot U ] ^2 $ are irrelevant. Similarly the terms
 in $A_{0,R}(x)$ are all irrelevant.

Next we have to consider the terms of the first order expansion in
external variables in (\ref{exte}), for which we need to develop the
$f$ function only to second order. We have
\begin{eqnarray}
A_{1}= \int dx dy dx^{0}\, \bar\varphi^{\les
e}(x,x^{0})\varphi^{\les
e}(y,x^{0})\Big[\mathfrak{U}\cdot\nabla\delta(x-y)\Big]\nonumber
e^{\imath (URU+ USW)}\prod_{l \in G^{i}_{k} \; , l \not \in T}
 du_{l}  d w_{l}\nonumber\\
\times C_{l}(u_l, w_l)\prod_{\ell\in T^{i}_{k} } du_{\ell}
du^{0}_{\ell} C_{\ell}(u_\ell,u^{0}_\ell, U_\ell, W_\ell) \big(
f(0)+f'(0)+\int_0^1 dt (1-t) f''(t) dt \big)\nonumber\\
&&
\end{eqnarray}

The first term is
\begin{eqnarray}
A_{1,0} &=& \int dx dy dx^{0}\, \bar\varphi^{\les
e}(x,x^{0})\varphi^{\les
e}(y,x^{0})\Big[\mathfrak{U}\cdot\nabla\delta(x-y)\Big] e^{\imath
(URU+ USW)}\nonumber\\&\times&\prod_{l \in G^{i}_{k} \; , l \not \in
T}
 du_{l}  d w_{l}
\times C_{l}(u_l, w_l)\prod_{\ell\in T^{i}_{k} } du_{\ell}
du^{0}_{\ell} C_{\ell}(u_\ell,u^{0}_\ell, U_\ell, W_\ell),
\end{eqnarray}
which vanishes identically due to the parity.

The second term is
\begin{eqnarray}
A_{1,0} &=& \int dx dy dx^{0}\, \bar\varphi^{\les
e}(x,x^{0})\varphi^{\les
e}(y,x^{0})\Big[\mathfrak{U}\cdot\nabla\delta(x-y)\Big] e^{\imath
(URU+ USW)}\nonumber\\&\times&\prod_{l \in G^{i}_{k} \; , l \not \in
T}
 du_{l}  d w_{l}
\times C_{l}(u_l, w_l)\prod_{\ell\in T^{i}_{k} } du_{\ell}
du^{0}_{\ell} C_{\ell}(u_\ell,u^{0}_\ell, U_\ell, W_\ell)\nonumber\\
&\times&[iXQU-2\mathcal {A}X(W+U)] .
\end{eqnarray}

The first term in the r.h.s. is
\begin{eqnarray}\label{2point3}
&&\int dx dx^{0}\, \bar\varphi^{\les
e}(x,x^{0})\mathfrak{U}\cdot(-i) \nabla\varphi^{\les e}(x,x^{0})
e^{\imath (URU+ USW)}\nonumber\\&\times&\prod_{l \in G^{i}_{k} \; ,
l \not \in T}
 du_{l}  d w_{l}
\times C_{l}(u_l, w_l)\prod_{\ell\in T^{i}_{k} } du_{\ell}
du^{0}_{\ell} C_{\ell}(u_\ell,u^{0}_\ell, U_\ell, W_\ell) \times XQU
\end{eqnarray}
It is logarithmically divergent and is proportional to the term
$\bar\varphi(x\wedge p)\varphi$ \footnote{We thank our referee for
pointing out this important point.}. But this term is also
vanishing, as for each graph there is always a mirror graph (
see\cite{DisertoriRivasseau2006} for details ) that is the same as
the former but reflected in a mirror. When we add them up, e.g. the
tadpole up with the tadpole down, the result is zero. \footnote{We
are very grateful to Prof. Rivasseau for explaining this.}

\begin{itemize}
 \item The term $\mathcal {A}XW\cdot\mathfrak{U}\cdot\nabla\varphi^{\les e}(x,x^{0}) $. The operator $\nabla$ brings a factor
 $M^{e}$,$\mathcal {A}$ brings a factor $M^{-2i(\ell)}$,
 $\mathfrak{U}$ brings a factor $M^{-i}$ and $W$ brings
 $M^{i(l)}$. The final factor is $M^{-(i+i(\ell)-2e)}\cdot M^{-(i(\ell)-i(l))}$.
 We remark that the scale of a tree line is higher than the
 loop line that lies over it, or the loop line would be chosen as
 tree line instead. So $i(\ell) > i(l)$ and this term is irrelavent.

\item The term $\mathcal {A}XU\cdot\mathfrak{U}\cdot\nabla\varphi^{\les e}(x,x^{0}) $ is smaller as $U$ brings
$M^{-i}$ and there is no long loop variables. So it is irrelavent.

\item We can easily find that $A_{1,R}$ is smaller hence irrelavent.
\end{itemize}

Now we consider the second order expansion in external variables in
(\ref{exte}). We only have to expand $f(t)$ to first order. We have
\begin{eqnarray}
A_{2}&=& \int dx dy dx^{0}\, \bar\varphi^{\les
e}(x,x^{0})\varphi^{\les
e}(y,x^{0})\frac{1}{2}\Big[(\mathfrak{U}\cdot\nabla)^{2}\delta(x-y
)\Big]\nonumber e^{\imath (URU+ USW)}\nonumber\\
&\times&\prod_{l \in G^{i}_{k} \; , l \not \in T}
 du_{l}  d w_{l} C_{l}(u_l, w_l)\prod_{\ell\in T^{i}_{k} } du_{\ell}
du^{0}_{\ell} C_{\ell}(u_\ell,u^{0}_\ell, U_\ell, W_\ell)\nonumber\\
&\times& \big( f(0)+\int_0^1 dt f'(t) dt \big).
\end{eqnarray}
The first term is
\begin{eqnarray}
A_{2,0}&=& \int dx dy dx^{0}\, \bar\varphi^{\les
e}(x,x^{0})\varphi^{\les
e}(y,x^{0})\frac{1}{2}\Big[(\mathfrak{U}\cdot\nabla)^{2}\delta(x-y
)\Big]\nonumber e^{\imath (URU+ USW)}\nonumber\\
&\times&\prod_{l \in G^{i}_{k} \; , l \not \in T}
 du_{l}  d w_{l} C_{l}(u_l, w_l)\prod_{\ell\in T^{i}_{k} } du_{\ell}
du^{0}_{\ell} C_{\ell}(u_\ell,u^{0}_\ell, U_\ell, W_\ell).
\end{eqnarray}
The terms with $\mu \ne \nu$ do not survive by parity. The other
ones reconstruct a counterterm proportional to the Laplacian. The
power-counting of this factor $A_{2,0}$ is improved (with respect to
$A$) by a factor $M^{-2(i-e)}$ which makes it only logarithmically
divergent, as should be for a wave-function counterterm.

The second term is
\begin{eqnarray}
A_{2,0}&=& \int dx dy dx^{0}\, \bar\varphi^{\les
e}(x,x^{0})\varphi^{\les
e}(y,x^{0})\frac{1}{2}\Big[(\mathfrak{U}\cdot\nabla)^{2}\delta(x-y
)\Big]\nonumber e^{\imath (URU+ USW)}\nonumber\\
&\times&\prod_{l \in G^{i}_{k} \; , l \not \in T}
 du_{l}  d w_{l} C_{l}(u_l, w_l)\prod_{\ell\in T^{i}_{k} } du_{\ell}
du^{0}_{\ell} C_{\ell}(u_\ell,u^{0}_\ell, U_\ell, W_\ell)\nonumber\\
&\times&\int_{0}^{1} dt (iXQU-2tAXX-2AX(W+U)) .
\end{eqnarray}

It is irrelevant as the terms in the integral bring at least a
convergent factor $M^{-(i-e)}$.

Putting together the results of the two previous section, we have
proved that the usual effective series  which expresses any
connected function of the theory in terms of an infinite set of
effective couplings, related one to each other by a discretized flow
\cite{Riv1}, have finite coefficients to all orders. Reexpressing
these effective series in terms of the renormalized couplings would
reintroduce in the usual way the Zimmermann's forests of
counterterms and build the standard renormalized series. The most
explicit way to check finiteness of these renormalized series in
order to complete the ``BPHZ theorem" is to use the ``classification
of forests"  which distributes Zimmermann's forests into packets
such that the sum over assignments in each packet is finite
\cite{Riv1}. This part is  identical to the
commutative case. Hence the proof of Theorem \ref{BPHZ1} is
completed.

\appendix
\section{The non-commutative $\varphi^6_3$ Model}
\setcounter{equation}{0}

In this appendix we discuss briefly the non-orientable real scalar
$\varphi^6_3$ model, and its renormalizability which is
questionable.

The action functional, with the notations of (\ref{action}) is now
\begin{eqnarray}\label{actionapp}
 S[\varphi] &=& \int \,d^2 x\ d x^0 \Big( \frac{1}{2}
 \partial_\mu \varphi \star
 \partial^\mu \varphi + \frac{1}{2} \partial_0 \varphi \star\partial^0 \varphi + \frac{\Omega^2}{2} (\tilde{x}_\mu \varphi )
 \star (\tilde{x}^\mu \varphi ) + \frac{1}{2} \mu_0^2 \,\varphi
 \star \varphi \nonumber\\
 &+& \frac{\lambda}{4} \varphi \star \varphi \star \varphi \star
 \varphi + \frac{g}{6} \varphi \star \varphi \star \varphi \star
 \varphi \star \varphi \star \varphi \Big)(x)\ .
\end{eqnarray}

In the real $\varphi^6$ model, there are only two kinds of
cyclically invariant vertices, namely the $\varphi^4 $ term:
\begin{eqnarray}\label{vertexphi4}
V(x_1, x_2, x_3, x_4) &=& \delta(x_1 -x_2+x_3-x_4)e^{i \sum_{1 \le
i<j \le 4}(-1)^{i+j+1}x_i \theta^{-1}  x_j}
\end{eqnarray}
and the $\varphi^6$ term:
\begin{eqnarray}\label{vertex6app}
 V(x_1, x_2, x_3, x_4, x_5, x_6) &=& \delta(x_1
 -x_2+x_3-x_4+x_5-x_6) \nonumber \\
 && e^{i
 \sum_{1 \le i<j \le 6}(-1)^{i+j+1}x_i \theta^{-1}x_j}
\end{eqnarray}
times the local factor in the time direction.
Again we note  $x \theta^{-1}  y  \equiv  \frac{2}{\theta} (x_1 y_2
-  x_2  y_1 )$.

The discussion is almost the same as that in
$(\bar\varphi\star\varphi)^3$ model, with a difference in the power
counting of the non-orientable graph.

When several disjoint $G^i_k$ subgraphs are non-orientable it is
better to solve longer clashing loop variables, essentially one per
disjoint non-orientable $G^i_k$, because they spare higher costs
than if tree lines were chosen instead. We define $S$ to be the set
of $n$ long variables to be solved via the $\delta$ functions. First
we put in $S$ all the $n-1$ long tree variables $v_l$. Then we scan
all the connected components $G^i_k$ starting from the leaves
towards the root, and we add a clashing line to $S$ each time when a
new non-orientable component $G^i_k$ appears. We also remove $p-1$
tree lines from $S$ so that each time $p\ges 2$ non-orientable
components merge into a single one. In the end we obtain a new set
$S$ of exactly $n-1+p-(p-1)=n$ long variables. So thanks to
inductive use of Lemma \ref{lemmarouting} in each $G^i_k$, we can
solve all the long variables in the set $S$ with the branch system
of $\delta$ functions associated to $T$ plus an additional loop
variable. But for the commutative dimension, there are always $n-1$
short tree variables to be integrated. So for a general non-orientable
graph we will earn only a convergent factor $M^{-2i}$ and the degree of
divergence given by this crude analysis becomes
$\omega(G^i_k)=N(G^i_k)/2+n_4-1$  (recall lemma \ref{crudelemma}).

Let us consider the improved analysis taking oscillations into
account. From the analog of lemma \ref{improvedbound} we see that
graphs with $g=0$, $n_4=0$, $B=2$ and $N=2$ remain dangerous. Such
graphs can't appear in the $(\bar\varphi\star\varphi)^3$ model as
they are non-orientable. In the $\varphi^{\star 6}$ model they can
appear, are logarithmic divergent and don't look like the initial
quadratic terms in the Lagrangian. So the two point function of this
$(\bar\varphi\star\varphi)^3$ model  seems non-renormalizable, but
maybe the situation can be rescued by combining all renormalizations
together, as is done e.g. in \cite{RenNCGN05} or maybe we can solve
this problem by exploring further the vertex oscillations. The study
of this problem is still in progress.

\section*{Acknowledgement}

The author wzht is very grateful to V. Rivasseau for successful
direction and encouragement and many useful  discussions on this
work. We are also very grateful to the anonymous referee for critical
comments. This work is supported by NSFC Grant No.10675108.

\providecommand{\href}[2]{#2}\begingroup\raggedright\endgroup


\begin{thebibliography}{10}

\bibitem{a.connes98:noncom}
A.~Connes, M.~R. Douglas, and A.~Schwarz, ``non-commutative geometry
and matrix
  theory: Compactification on tori,'' {\em JHEP} {\bf 02} (1998) 003,
  \href{http://www.arXiv.org/abs/hep-th/9711162}{{\tt hep-th/9711162}}.

\bibitem{Seiberg:1999vs}
N.~Seiberg and E.~Witten, ``String theory and non-commutative
geometry,'' {\em
  JHEP} {\bf 09} (1999) 032,
\href{http://www.arXiv.org/abs/hep-th/9908142}{{\tt
hep-th/9908142}}.
%%CITATION = HEP-TH 9908142;%%.


\bibitem{DouNe}
M.~R. Douglas and N.~A. Nekrasov, ``non-commutative field theory,''
{\em Rev.
  Mod. Phys.} {\bf 73} (2001) 977--1029,
\href{http://www.arXiv.org/abs/hep-th/0106048}{{\tt
hep-th/0106048}}.
%%CITATION = HEP-TH 0106048;%%.

\bibitem{szab}
R.~J. Szabo `` Quantum field
  theory on non-commutative phase spaces,'' {\em Phys.Rep.} {\bf 378} (2003) 207,
\href{http://www.arXiv.org/abs/hep-th/0109162}{{\tt
hep-th/0109162}}.
%%CITATION = HEP-TH 0109162;%%.


\bibitem{Minw}
S.Minwalla,M. Van Raamsdonk,and N.Seiberg, ``Nocommutative
perturbative dynamics,'' {\em JHEP}
  {\bf 02} (2000), 020,
\href{http://www.arXiv.org/abs/hep-th/9912072}{{\tt
hep-th/9912072}}.
%%CITATION = HEP-TH 9912072;%%.


\bibitem{GrWu03-1}
H.~Grosse and R.~Wulkenhaar, ``Power-counting theorem for non-local
matrix  models and renormalisation,'' {\em Commun. Math. Phys.} {\bf
254} (2005),  no.~1, 91--127,
\href{http://www.arXiv.org/abs/hep-th/0305066}{{\tt
hep-th/0305066}}.
%%CITATION = HEP-TH 0305066;%%.

\bibitem{GrWu04-3}
H.~Grosse and R.~Wulkenhaar, ``Renormalisation of
$\varphi^{4}$-theory on  non-commutative $\mathbb{R}^4$ in the
matrix base,'' {\em Commun. Math. Phys.}  {\bf 256} (2005), no.~2,
305--374, \href{http://www.arXiv.org/abs/hep-th/0401128}{{\tt
hep-th/0401128}}.
%%CITATION = HEP-TH 0401128;%%.

\bibitem{LaSz}
E.~Langmann and R.~J. Szabo, ``Duality in scalar field theory on
non-commutative
  phase spaces,'' {\em Phys. Lett.} {\bf B533} (2002) 168--177,
\href{http://www.arXiv.org/abs/hep-th/0202039}{{\tt
hep-th/0202039}}.
%%CITATION = HEP-TH 0202039;%%.

\bibitem{Polch}
J.~Polchinski, Renormalization and Effective Lagrangians, {\em Nucl.
Phys.} {\bf B231} (1984) 269.

\bibitem{Rivasseau2005bh}
V.~Rivasseau, F.~Vignes-Tourneret, and R.~Wulkenhaar,
``Renormalization of
  non-commutative $\varphi^4$-theory by multi-scale analysis,'' {\em Commun. Math.
  Phys. (Online First) DOI: 10.1007/s00220-005-1440-4} (2005)
\href{http://www.arXiv.org/abs/hep-th/0501036}{{\tt
hep-th/0501036}}.
%%CITATION = HEP-TH 0501036;%%.

\bibitem{xphi4-05}
R.~Gurau, J.~Magnen, V.~Rivasseau and F.~Vignes-Tourneret,
Renormalization of non-commutative $\varphi^4_4$ field theory in $x$
space, {\em Commun.~Math.~Phys.} {\bf 267} (2006), no.~2, 515--542,
\href{http://www.arXiv.org/abs/hep-th/0512271}{{\tt
hep-th/0512271}}.


\bibitem{gurauhypersyman}
R.~Gurau and V.~Rivasseau, Parametric representation of
non-commutative field
  theory, to appear in Commun.~Math.~Phys, \href{http://www.arXiv.org/abs/math-ph/0606030}{{\tt math-
ph/0606030}}.

\bibitem{GrWu04-2}
H.~Grosse and R.~Wulkenhaar, The beta-function in duality-covariant
  non-commutative $\varphi^{4}$-theory, {\em Eur. Phys. J.} {\bf C35} (2004)
  277--282,
\href{http://www.arXiv.org/abs/hep-th/0402093}{{\tt
hep-th/0402093}}.
%%CITATION = HEP-TH 0402093;%%.

\bibitem{DisertoriRivasseau2006}
M.~Disertori and V.~Rivasseau, Two and Three Loops Beta Function of
Non Commutative
 $\Phi^4_4$ Theory \href{http://www.arXiv.org/abs/hep-th/0610224}{{\tt hep-th/0610224}}.
%%CITATION = HEP-TH 0610224;%%

\bibitem{beta2-06}
M. Disertori, R.~Gurau, J.~Magnen and V.~Rivasseau, Vanishing of
Beta Function of Non Commutative $\Phi_4^4$ to all orders, Submitted
to Phys.~Lett.~B,
\href{http://www.arXiv.org/abs/hep-th/0612251}{{\tt
hep-th/0612251}}.


\bibitem{RenNCGN05}
F.~Vignes-Tourneret, Renormalization of the orientable
non-commutative  {Gross-Neveu} model. To appear in
Ann.~H.~Poincar\'e, \href{http://www.arXiv.org/abs/math-ph/
0606069}{{\tt math-ph/0606069}}.


\bibitem{Langmann:2003if}
E.~Langmann, R.~J. Szabo, and K.~Zarembo, ``Exact solution of
quantum field
  theory on non-commutative phase spaces,'' {\em JHEP} {\bf 01} (2004) 017,
\href{http://www.arXiv.org/abs/hep-th/0308043}{{\tt
hep-th/0308043}}.
%%CITATION = HEP-TH 0308043;%%.


\bibitem{Grosse2005ig}
H.~Grosse and H.~Steinacker, Renormalization of the non-commutative
$\varphi^3$
  model through the kontsevich model. 2005.

\bibitem{Grosse2006qv}
H.~Grosse and H.~Steinacker, A nontrivial solvable non-commutative
$\varphi^3$
  model in $4$ dimensions,
\href{http://www.arXiv.org/abs/hep-th/0603052}{{\tt
hep-th/0603052}}.
%%CITATION = HEP-TH 0603052;%%.

\bibitem{Grosse2006er}
H.~Grosse and H.~Steinacker, Exact renormalization of a
non-commutative phi**3 model in 6 dimensions,
\href{http://www.arXiv.org/abs/hep-th/hep-th/0607235}{{\tt
hep-th/0607235}}.
%%CITATION = HEP-TH hep-th/0607235;%%.


\bibitem{rivfvt}
Vincent Rivasseau and Fabien Vignes-Tourneret, ``Renormalisation of
non-commutative field theories''
\href{http://www.arXiv.org/abs/hep-th/0702068}{{\tt
hep-th/0702068}}.
%%CITATION = HEP-TH 0702068;%%.

\bibitem{riv07}
Vincent Rivasseau,``Non-Commutative Renormalization''
\href{http://lanl.arxiv.org/abs/0705.0705}{{\tt hep-th/0705.0705}}.
%%CITATION = HEP-TH 0705.0705.;%%.

\bibitem{Suss} L. Susskind, The Quantum Hall Fluid and Non-Commutative Chern Simons Theory,
hep-th/0101029

\bibitem{Poly} A. Polychronakos,
Quantum Hall states as matrix Chern-Simons theory, JHEP 0104 (2001)
011, \href{http://www.arXiv.org/abs/hep-th/0103013}{{\tt
hep-th/0103013}}

\bibitem{HellRaam} S. Hellerman and M. Van Raamsdonk,
Quantum Hall Physics = non-commutative Field Theory, JHEP 0110 (2001)
039, \href{http://www.arXiv.org/abs/hep-th/0103179}{{\tt hep-th/
0103179}}



\bibitem{toolbox05}
R.~Gurau, V.~Rivasseau, and F.~Vignes-Tourneret, ``Propagators for
  non-commutative field theories,''
  \href{http://www.arXiv.org/abs/hep-th/0512071}{{\tt hep-th/0512071}}.
  submitted to Ann.~H.~Poincar\'e.


\bibitem{Riv1}
V.~Rivasseau, {\em From Perturbative to Constructive
Renormalization}.
\newblock Princeton series in physics. Princeton Univ. Pr., Princeton, USA,
  1991.
\newblock 336 p.


\bibitem{Filk:1996dm}
T.~Filk, ``Divergencies in a field theory on quantum space,'' {\em
Phys. Lett.}  {\bf B376} (1996) 53--58.
%%CITATION = PHLTA,B376,53;%%.

\bibitem{Chepelev:2000hm}
I.~Chepelev and R.~Roiban, ``Convergence theorem for non-commutative
feynman
  graphs and renormalization,'' {\em JHEP} {\bf 03} (2001) 001,
\href{http://www.arXiv.org/abs/hep-th/0008090}{{\tt
hep-th/0008090}}.
%%CITATION = HEP-TH 0008090;%%.

\bibitem{GaN}
G.~Gallavotti and F.~Nicol\`o, ``Renormalization theory in
four-dimensional
  scalar fields. i,'' {\em Commun. Math. Phys.} {\bf 100} (1985)
545--590.
%%CITATION = CMPHA,100,545;%%.




\end{thebibliography}
\end{document}